\documentclass[pra,twocolumn,superscriptaddress,amsmath]{revtex4-1}

\usepackage{graphicx,float}
\usepackage{braket}

\begin{document}
\title{Adiabatic Fock state generation scheme using Kerr nonlinearity}
\author{Ryotatsu Yanagimoto}
\email[Email: ]{ryotatsu@stanford.edu}
\affiliation{Edward L.\ Ginzton Laboratory, Stanford University, Stanford, CA 94305, USA}
\author{Edwin Ng}
\affiliation{Edward L.\ Ginzton Laboratory, Stanford University, Stanford, CA 94305, USA}
\author{Tatsuhiro Onodera}
\affiliation{Edward L.\ Ginzton Laboratory, Stanford University, Stanford, CA 94305, USA}
\author{Hideo Mabuchi}
\affiliation{Edward L.\ Ginzton Laboratory, Stanford University, Stanford, CA 94305, USA}

\begin{abstract}
We propose a theoretical scheme to deterministically generate Fock states in a Kerr cavity thorough adiabatic variation of the driving field strength and the cavity detuning. We show that the required time to generate a $n$-photon Fock state scales as the square root of $n$. Requirements for the Kerr coefficient relative to the system decoherence rate are provided as a function of desired state fidelity, indicating that the scheme is potentially realizable with the present state of the art in microwave superconducting circuits.  
\end{abstract}

\maketitle

Generation and manipulation of non-classical photon states are important techniques in quantum engineering. Among these non-classical states, Fock states, which have deterministic photon number, are of particular importance. However, despite being conceptually simple, the generation of Fock states tends to be nontrivial as it requires some source of strong optical nonlinearity.

Several theoretical schemes employ quantum non-demolition (QND) measurements of the field inside a cavity in order to collapse the state to a Fock state \cite{Brune1990,Holland1991} or even to an arbitrary quantum state \cite{Vogel1993}. Experimental realization of Fock states by means of QND have been reported \cite{Guerlin2007,Waks2006}. It is worth mentioning that these schemes are conditioned upon the result of the measurements. 

On the other hand, schemes based on unitary evolution of known initial states are capable of generating desired quantum states deterministically. Theoretical schemes based on nonlinearities induced by atoms or qubits have been proposed for the deterministic manipulation of photonic states \cite{Brown2003,Parkins1993,Parkins1995,Law1996,Lange2000}, and the recent progress in circuit quantum electrodynamics has enabled the experimental realization of these schemes \cite{Wang2008,Hofheinz2008,Hofheinz2009,Premaratne2017}.

The Kerr effect is another possible means to overcome the intrinsic harmonicity of photonic systems. Using Kerr-based nonlinear Mach-Zehnder interferometers, it has been shown that the photon number uncertainty of the output light can be reduced \cite{Kitagawa1986,Sundar1996}. Each of Ref.~\cite{Kuklinski1990} and Ref.~\cite{Leonski1997} shows that Fock states can be formed inside a Kerr nonlinear cavity driven by periodic pulses through different processes respectively. Though sufficient nonlinearity for these schemes is not easy to reach, single-photon Kerr nonlinearities have been observed recently in the microwave regime \cite{Kirchmair2013}.

In this work, we propose a new scheme to deterministically generate a Fock state inside a Kerr nonlinear cavity thorough adiabatic variation of experimental parameters: detuning of the cavity and the strength of the field driving the cavity. Because of its adiabatic nature, our method has the advantages of being timing insensitive and robust to the noise in the driving field. Also, as this scheme utilizes the Kerr effect, it can potentially be implemented on chip in the optical regime, considering the recent progress in research of monolayer materials with high optical Kerr coefficients \cite{Soh2018}. We provide a formalism to optimize the adiabatic sequence to generate a Fock state. Based on numerical simulations, we argue that the time required to generate a $n$-photon Fock state scales as $\sqrt{n}$, which is beneficial when large Fock states are desired. In addition, we provide estimates for the Kerr coefficient required to achieve a particular fidelity, relative to the system decoherence rate in the presence of linear loss. It turns out that experimental demonstration of our scheme is realistic using superconducting circuit architectures~\cite{Peterer2015}.

We first introduce our model of the system. We consider a Kerr nonlinear cavity with resonant frequency $\omega_\mathrm{c}$ driven by a field with a frequency $\omega_\mathrm{d}$. The Hamiltonian of the system in the rotating frame of the driving field is
\begin{equation}\label{hamiltonian}
H=\frac{\chi}{2}{a^\dagger}^2a^2+\Delta a^\dagger a+\beta\left(a+a^\dagger\right),
\end{equation} 
where $a$ denotes the annihilation operator of photons inside the cavity, $\chi$ is the Kerr coefficient, and $\Delta=\omega_\mathrm{c}-\omega_\mathrm{d}$ is the cavity detuning. Here, $\beta$ denotes the strength of the cavity driving field, and  we assume $\beta$ to be a non-negative real number without loss of generality. In the following, we set $\chi=1$, which sets the time scale. 

\begin{figure}[h]
	\includegraphics[width=0.5\textwidth]{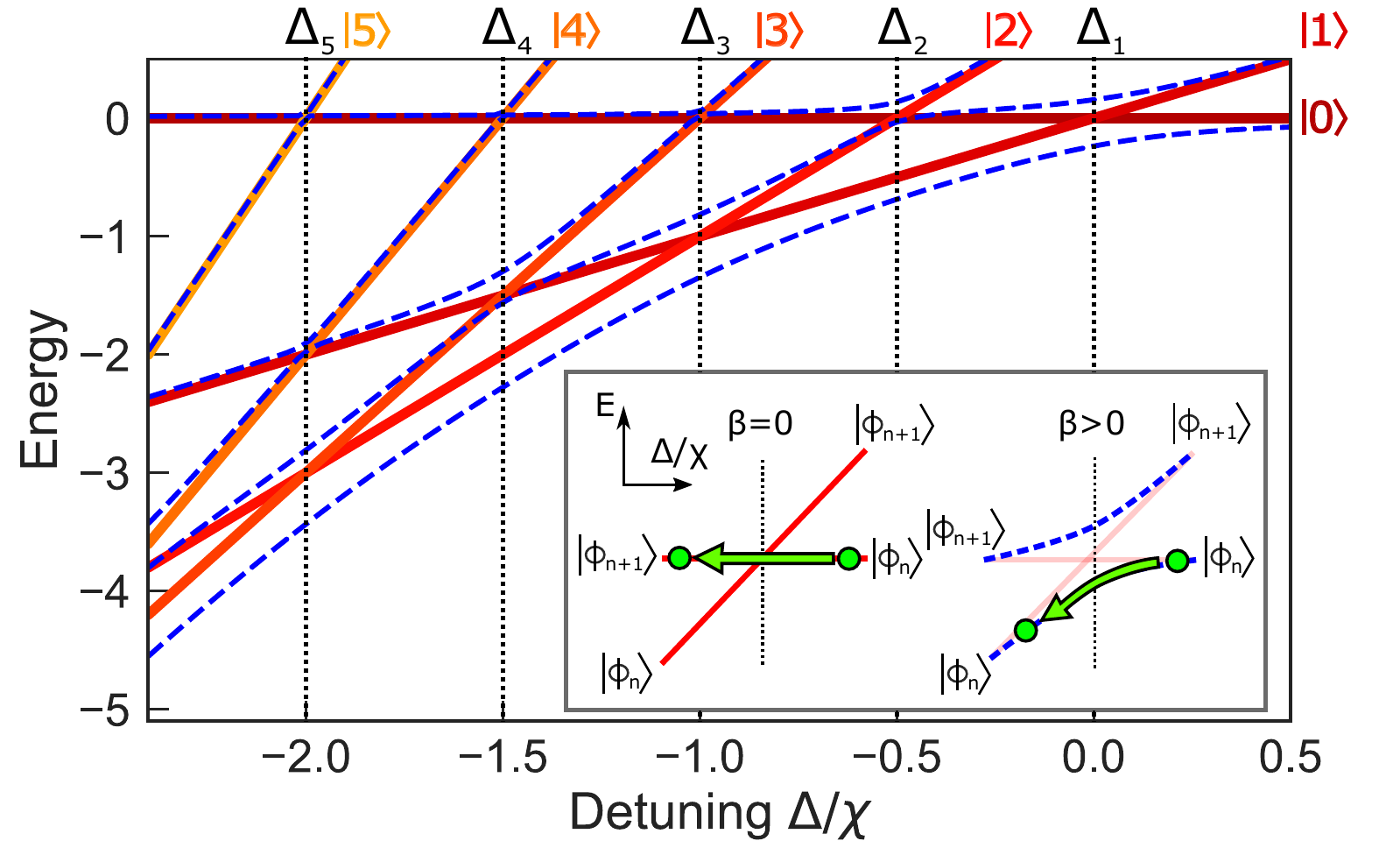}

	\caption{Eigenenergies of the Hamiltonian \eqref{hamiltonian} plotted as functions of detuning $\Delta$ for drives $\beta=0$ (solid lines labeled with $\ket{n}$) and $\beta=0.2$ (blue dashed lines). Vertical black dotted lines indicate the detunings $\Delta_l$ corresponding to energy crossing points. Inset: When detuning is adiabatically varied over an energy crossing point, the state can either cross over to the neighboring energy level when $\beta=0$ (left), or it can stay on the same energy level when $\beta>0$ (right). }
		\label{energy}
\end{figure}

For the Hamiltonian \eqref{hamiltonian}, we denote the $k^\text{th}$ energy eigenstate at a given $\Delta$ and $\beta$ as $\ket{\phi_k(\Delta,\beta)}$ with eigenenergy $E_k$. In the absence of the drive $(\beta=0)$, eigenstates of the Hamiltonian are Fock states $\ket{n}$ with eigenenergies $\frac{1}{2}n\left(n-1\right)+n\Delta$. As shown in Fig.~\ref{energy}, two such eigenstates $\ket{n}$ and $\ket{m}$ are degenerate at the detuning $\Delta_{n+m}$, where we define  $\Delta_l\equiv-\left(l-1\right)/2$.
More specifically, the eigenstates of the system with no drive are
\begin{equation}\label{nopumpstate}
\begin{split}
&\ket{\phi_k(\Delta,0)}=
\begin{cases}
\ket{k} & \Delta\in[\Delta_k,\infty)\\
\ket{n} & \Delta\in[\Delta_{k+2n+1},\Delta_{k+2n})\\
\ket{k+n} & \Delta\in[\Delta_{k+2n},\Delta_{k+2n-1})
\end{cases}.
\end{split}
\end{equation}

In the presence of a finite drive $\beta>0$, all degeneracies are lifted as shown in Fig.~\ref{energy}. As can be seen from the figure, the energy gaps between the ground states $\ket{\phi_0(\Delta,\beta)}$ and the first excited states $\ket{\phi_1(\Delta,\beta)}$ are significantly larger than the other gaps. This is because, for $\beta\rightarrow0$, $\ket{\phi_0(\Delta,\beta)}$ and $\ket{\phi_1(\Delta,\beta)}$ converge to neighboring Fock states which interact at first order in $\beta(a+a^\dagger)$, while the interactions between other states are mediated by higher order terms. These relatively larger energy gaps play an important role in our generation scheme.

The effect of $\beta$ on the presence of the energy gaps enables the control of the system state when combined with adiabatic control over $\Delta$. As shown schematically in the inset of Fig.~\ref{energy}, when the detuning is varied over an energy crossing point with $\beta=0$, the system state is transfered to the neighboring energy state. On the other hand, with a finite drive $\beta>0$, the system state remains in the same energy eigenstate over the energy crossing point provided that detuning $\Delta$ varies sufficiently slowly. 

Our Fock state generation scheme utilizes the above observations as follows. We initialize the system in the vacuum state $\ket{0}$ with $\beta=0$ and proceed to adiabatically vary $\Delta$ and $\beta$ to arrive at a $n$-photon Fock state $\ket{n}$. In order to take advantage of the first-order gaps between the ground and first-excited states, we would like the state at all times to be on the ground state; for the initial state, this is realized when the initial detuning is some $\Delta_\text{i}>0$. Furthermore, this also implies that the final values of the parameters are $\beta=0$ and $\Delta=\Delta_\text{f}\in\left(\Delta_{2n+1},\Delta_{2n-1}\right)$. Finally, for intermediate parameter values, $\beta$ has to be non-zero when $\Delta$ passes $\Delta_{1}, \Delta_{3},\dots,\Delta_{2n-1}$ to maintain the non-zero gaps at those points. Under these conditions, and provided that the parameter variations are sufficiently slow, the final state will be the desired Fock state by the adiabatic theorem.

To determine the exact shape of the path, we first parametrize the path as $C=\left\{\left(\Delta(s),\beta(s)\right): s\in\left[0,S\right]\right\}$ where $\mathrm{d}s=\mathrm{d}\sqrt{\Delta^2+\beta^2}$ is the line element in $(\Delta,\beta)$-space, and $S$ is the total length of the path. The trajectories of the parameters are obtained by introducing time dependence in $s$ so that $s(t)$ represents the distance covered up to time $t$. From the condition for adiabaticity, an instantaneous penalty function $P_C(t)$ characterizing the probability of leaving the ground state is
\begin{equation}
\begin{split}
P_C(t)=\sum_{n\ne 0}\frac{\left\vert\left<\phi_n(t)\middle|\frac{\mathrm{d} H}{\mathrm{d} t}\middle|\phi_0(t)\right>\right\vert}{\left\vert E_n(t)-E_0(t)\right\vert^2}=\frac{\mathrm{d}s}{\mathrm{d} t}\times Q_C(s).
\end{split}
\end{equation}
Note that the time dependencies of the parameters above are given via the time dependence of $s(t)$. Here, the function $Q_C$ is given by
\begin{equation}\label{qequation}
Q_C(s)=\sum_{n\ne 0}\frac{\left\vert \frac{\mathrm{d}\beta}{\mathrm{d} s}L_n(s)+\frac{\mathrm{d}\Delta}{\mathrm{d} s}M_n(s)\right\vert}{\left\vert E_n(s)-E_0(s)\right\vert^2},
\end{equation}
with $\ket{\phi_n(s)}=\ket{\phi_n(\Delta(s),\beta(s)}$ and
\begin{align}
L_n(s)&=\left<\phi_n(s)\middle|a^\dagger+ a\middle|\phi_0(s)\right\rangle\\
M_n(s)&=\left<\phi_n(s)\middle|a^\dagger a\middle|\phi_0(s)\right>.
\end{align}

The total penalty of the path $C$ is then calculated as
\begin{equation}\label{int}
I[C]=\int_0^T\mathrm{d}t~P_C(t)=\int_{0}^{S}\mathrm{d}s~Q_C(s),
\end{equation}
where $T$ is the total time to traverse the path so that $s(T)=S$. By minimizing the second integral, we can determine the best path without knowing the time dependence of $s(t)$. 

Given an (optimal) path $C$, we choose the time dependence $s(t)$ as follows. First, note that $P$ is a non-negative function of time, so by \eqref{int},
\begin{equation}\label{inequality}
\max_t P_C(t)\ge \frac{I[C]}{T}.
\end{equation} 
It is reasonable to choose a parametrization such that $P_C(t)=I[C]/T$ for all $t$ so that the inequality is always saturated. Under this choice, we have
\begin{equation}
\label{time}
t(s)=\frac{T}{I[C]}\int^s_0\mathrm{d}s'~Q_C(s').
\end{equation}
By numerically inverting $t(s)$, we then get $s(t)$, which in turn produces the time-dependent trajectories of parameters $\beta(t)$ and $\Delta(t)$ given the total desired time $T$.

\begin{figure}[h]
    \includegraphics[width=0.48\textwidth]{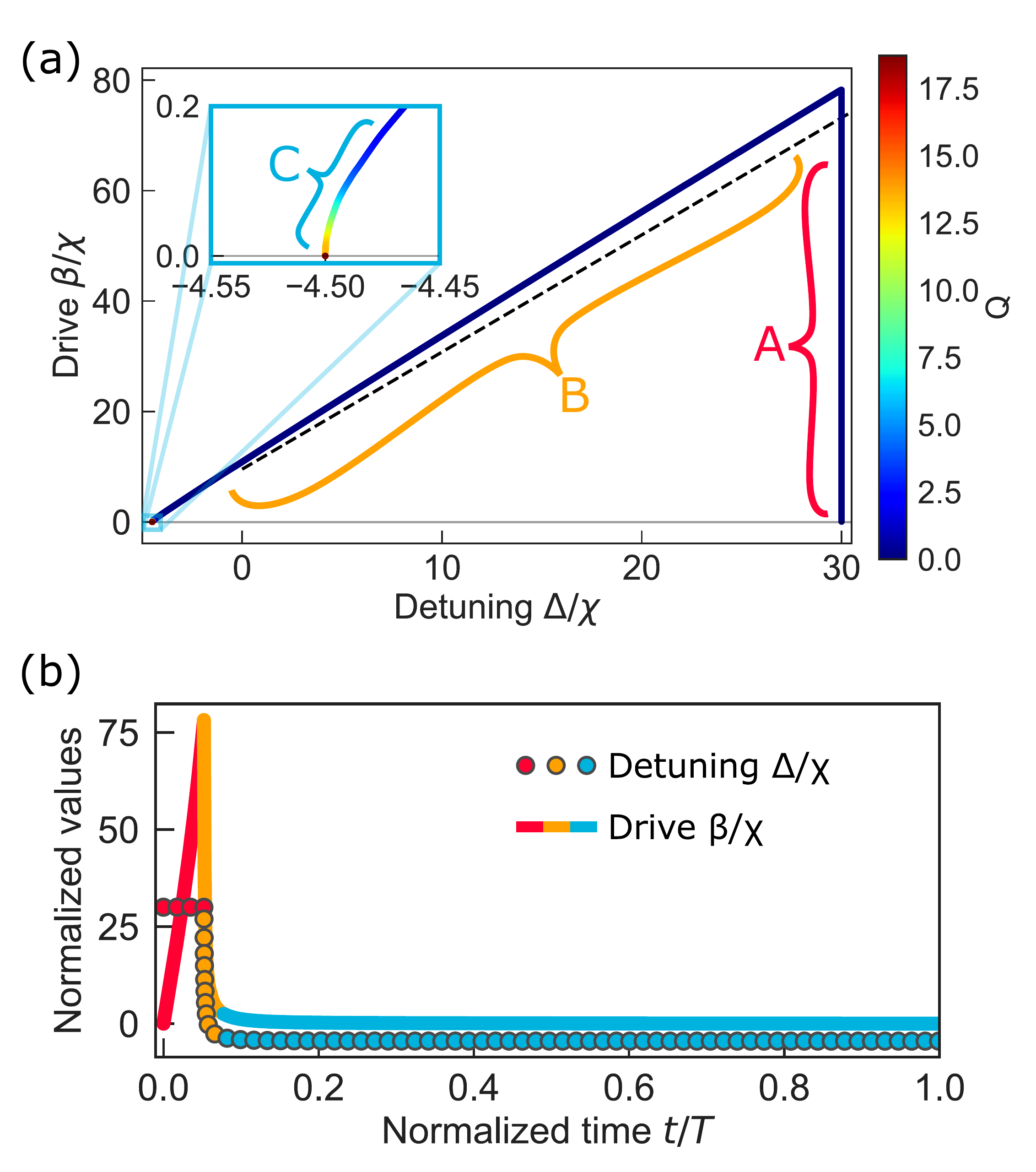}
    \caption{(a) Optimized path for generating $\ket{5}$ with final detuning $\Delta_\text{f}=-4.5$ and $\Delta_\text{max}=30$. The color of the path indicates the penalty $Q$ at each point. Regions A, B, and C are labeled as discussed in the text; here, C is defined to be where the penalty exceeds $10^{-4}$ its maximum. The inset shows a magnified image of the path around region C. The black dashed line indicates the approximation for region B based on \eqref{approxline}. (b) Drive $\beta(t)$ and detuning $\Delta(t)$ as prescribed by \eqref{time} plotted as functions of time. The colors represent to which region each point belongs.}
	\label{optimized}
\end{figure}

To numerically optimize the path, we approximate it with line edges connecting vertices. In an iterative procedure, the position of each vertex is perturbed. If the perturbation reduces the total penalty, we update the position of the vertex, and reject the perturbation otherwise. In Fig.~\ref{optimized}, we show the result of this optimization procedure for the generation of $\ket{5}$.

As a technical point, we observe that $\Delta_\text{i}$ tends to diverge, reflecting that the growth of the energy gaps $\left\vert E_n-E_0\right\vert\propto\Delta$ with increasing $\Delta_\text{i}$ dominates the penalty induced by the extended path length. In real experiments, however, arbitrarily large cavity detuning cannot be realized, so we set a bound on the detuning, $\Delta\le\Delta_\text{max}$. Though we set $\Delta_\text{max}=30$ for the following numerics, the choice of $\Delta_\text{max}$ does not have significant effects on the overall penalty, because it only alters the path where the penalty is relatively small. For the path shown in Fig.~\ref{optimized}, the reduction of the penalty achieved with $\Delta_\mathrm{max}\rightarrow\infty$ (vs.\ $\Delta_\mathrm{max}=30$) is only $2$\%.

As shown in Fig.~\ref{optimized}(a), the general structure of the optimized path is described by three qualitative regions named A, B, and C. Region A follows the initial increase of $\beta$ up to some $\beta_\mathrm{max}$ under fixed detuning $\Delta=\Delta_\mathrm{max}$. Then, in region B, the trajectory takes a diagonal path towards the terminal point $(\Delta_\mathrm{f},0)$. As it approaches the $\Delta$ axis, the path bends towards the vertical direction, which we denote as region C. As can be seen from the inset of Fig.~\ref{optimized}(a), the penalty rapidly increases as the path reaches the end of region C and takes its maximum value on the $\Delta$ axis. As region A is merely an artifact of upper-bounding $\Delta_\text{i}$, we focus our discussion on the structures of regions B and C, which will help explain both the achieved value $\beta_\mathrm{max}$ and our choice for $\Delta_\mathrm{f}$.
 
Numerically, we observe that the optimal curve for region B is approximately a straight line connecting two points $(\Delta_\mathrm{max},\beta_\mathrm{max})$ and $(\Delta_\mathrm{f},0)$, which can be understood by the following argument. For $\beta>0$ and $\Delta>0$, the ground state is well approximated by an ansatz coherent state $\mathcal{D}(\alpha_0)\ket{0}$, where $\mathcal{D}(\alpha)=\exp(\alpha a^\dagger-\alpha^*a)$ is the displacement operator. Using the variational method, $\alpha_0$ is given as the real solution of the equation
\begin{equation}\label{groundstate}
\alpha_0^3+\Delta\alpha_0+\beta=0.
\end{equation}
Furthermore, the first excited state is well approximated as $\mathcal{D}(\alpha_0)\ket{1}$, which leads to $L_1=1$ and $M_1=\alpha_0$. Limiting our interest to the coupling between the ground and first-excited states, the numerator of $Q$ vanishes when
\begin{equation}
\frac{\mathrm{d}\beta/\mathrm{d}s}{\mathrm{d}\Delta/\mathrm{d} s}=-M_1/L_1=-\alpha_0.
\end{equation}
Since the optimized path should fulfill this condition, it must approximately be a line with slope $\mathrm{d}\beta/\mathrm{d}\Delta=-\alpha_0$. Taking into account the terminal point $(\Delta_\mathrm{f},0)$, region B is therefore well approximated by a line
\begin{equation}\label{approxline}
\beta=\sqrt{-\Delta_\mathrm{f}}\left(\Delta-\Delta_\mathrm{f}\right),
\end{equation}
which is in good agreement with the numerical result as shown in Fig.~\ref{optimized}(a). Based on this result, we also see that $\beta_\mathrm{max}\approx\sqrt{-\Delta_\mathrm{f}}\left(\Delta_\mathrm{max}-\Delta_\mathrm{f}\right)$.

Next, we discuss the structure of region C. This region is of the highest importance, because the system spends most of its time in this region due to the high penalty, which reflects the drastic changes in the eigenstates. For $\beta\sim0$ and $\Delta\in\left(\Delta_{2n+1},\Delta_{2n-1}\right)$, we approximate the ground state as a displaced Fock state $\mathcal{D}(\alpha_n)\ket{n}$. Again $\alpha_n$ is variationally given as the real solution of
\begin{equation}
\alpha_n^3+(\Delta+2n)\alpha_n+\beta=0.
\end{equation}
Note that for $\beta\rightarrow0$, $\alpha_n$ goes to $0$, and thus the ansatz ground state $\mathcal{D}(\alpha_n)\ket{n}$ converges to the true ground state $\ket{n}$ in the absence of drive. Thus, the dominant cross-over states are $\mathcal{D}(\alpha_n)\ket{n-1}$ and $\mathcal{D}(\alpha_n)\ket{n+1}$. Inserting these two states into \eqref{qequation}, the leading-order contribution to the penalty in the vertical direction is
\begin{equation}\label{vpen}                                        Q_\beta=\left[\frac{\sqrt{n+1}}{\left(\frac{1}{2}+\delta\right)^2}+\frac{\sqrt{n}}{\left(-\frac{1}{2}+\delta\right)^2}\right]+\mathcal{O}(\beta^2),
\end{equation}
where $\delta=\Delta-(\Delta_{2n+1}+\Delta_{2n-1})/2$. As the penalty $Q_\Delta$ along the horizontal direction is zero up to $\mathcal{O}(\beta^2)$, it is expected that the path in region C is solely determined by $Q_\beta$. The absence of the first-order term in $\beta$ is intuitively understood through symmetry: as the choice of the drive phase is arbitrary, the penalty should not depend on the sign of $\beta$, and thus only even-order terms in $\beta$ appear.

Ignoring higher-order terms in $\beta$, the vertical penalty $Q_\beta$ is minimized when $\delta$ is given as the real solution of
\begin{equation}
\label{delta}
\left(\frac{\frac{1}{2}-\delta}{\frac{1}{2}+\delta}\right)^3=\sqrt{\frac{n}{n+1}},
\end{equation}
which is approximately
\begin{equation}
\label{approximate}
\delta\approx\frac{1}{6(\sqrt{n+1}+\sqrt{n})^2}\approx 0.
\end{equation}
Consider now a path approaching the $\Delta$ axis by traversing a horizontal narrow strip $0\le\beta\le\mathrm{d}\beta$. Using $Q_\Delta=0$, the minimum possible penalty upon traversing this strip to reach a point on the $\Delta$ axis with $\Delta\in\left(\Delta_{2n+1},\Delta_{2n-1}\right)$ is attained when the path is a vertical line with $\delta$ given by \eqref{delta}. These features are reproduced in region C of Fig.~\ref{optimized}(a) in the vicinity of the $\Delta$ axis. As a consequence of \eqref{approximate}, $\Delta_\mathrm{f}$ is forced to the midpoint of $(\Delta_{2n+1},\Delta_{2n-1})$ for sufficiently large $n$, which justifies our choice to fix $\Delta_\mathrm{f}=(\Delta_{2n+1}+\Delta_{2n-1})/2$ in the numerical optimization.

This latter analysis also allows us to estimate the scaling of our scheme. In order to adiabatically generate a state with a fixed fidelity, we need to bound the instantaneous penalty $P_C(t)$ throughout the sequence. By \eqref{inequality}, the time we need to spend on the sequence is proportional to the total penalty of the optimized path. As qualitatively shown in Fig.~\ref{optimized}, the total penalty is dominated by the contribution from region C, which in turn is dominated by the vertical penalty $Q_\beta$ given by \eqref{vpen}. Therefore, we conclude that the time $T_n$ required to generate $\ket{n}$ should scale as $\sqrt{n}$, or more specifically, $T_n\sim\sqrt{n}/\chi$.

Up to this point, the discussions have been based on a closed system analysis. In order to generalize to the open system case, we introduce a Lindblad operator $L=\sqrt{\kappa}a$ to describe single photon loss, and the dynamics of the system is described by the master equation
\begin{equation}
\label{master}
\dot{\rho}=-i\left[H,\rho\right]+L\rho L^\dagger-\frac{1}{2}L^\dagger L\rho-\frac{1}{2}\rho L^\dagger L. 
\end{equation} 

In the presence of single photon loss, a Fock state $\ket{n}$ decoheres on a time scale of $\tau_n\sim 1/n\kappa$ \cite{Lu1989,Wang2008}. The fidelity of the generated state suffers if the time spent on the sequence is long compared to $\tau_n$. As a result, requiring a fixed fidelity implies the existence of an upper bound for $T_n/\tau_n\sim n^{3/2}\kappa/\chi$. To meet this requirement, we therefore need at least $\chi/\kappa\sim n^{3/2}$.

\begin{figure}[h]
	\includegraphics[width=0.43\textwidth]{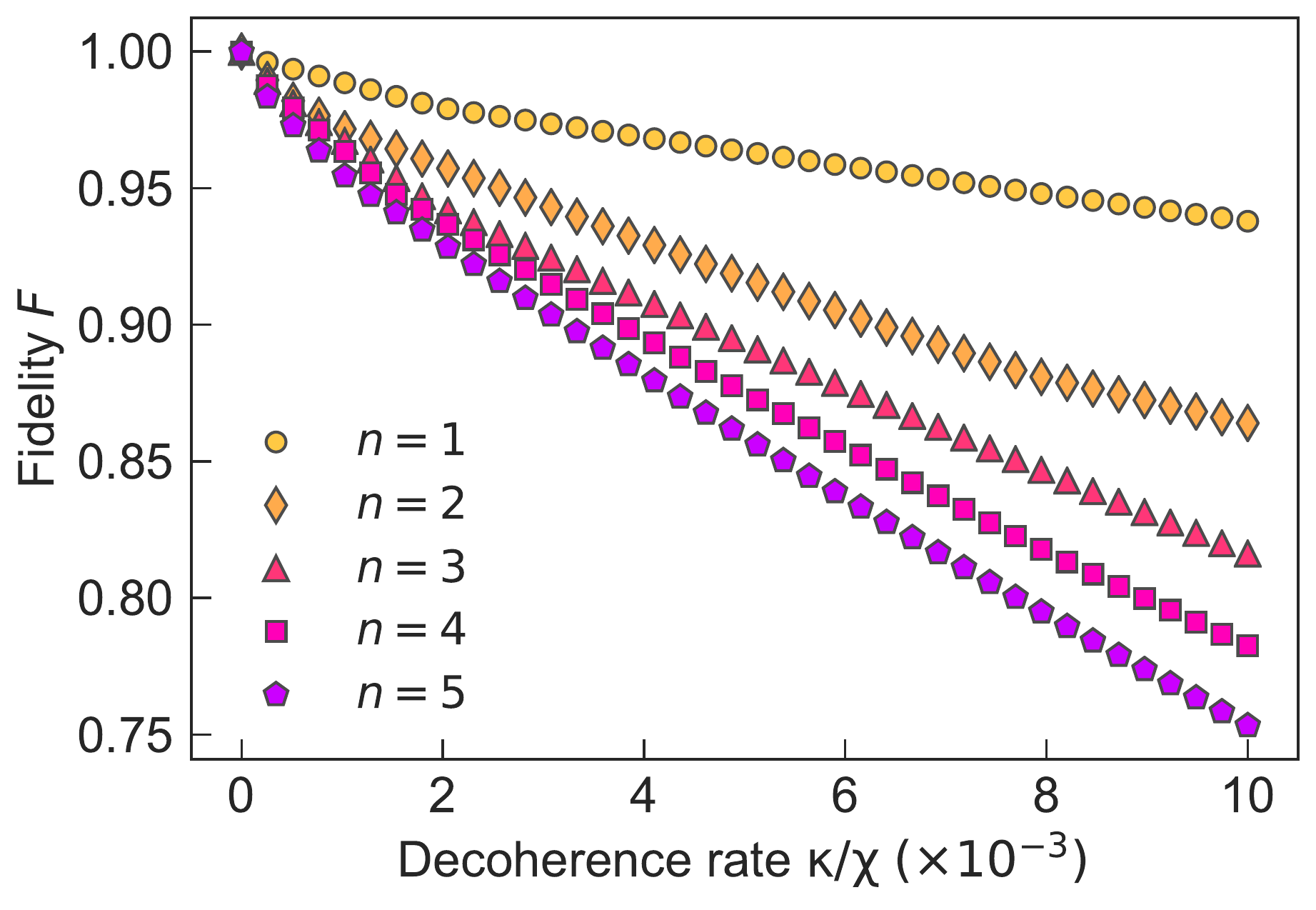}
	\caption{Highest achieved fidelity $F=\left<n\middle|\rho(T)\middle| n\right>$ for various target Fock states $\ket{n}$ as a function of decoherence rate $\kappa/\chi$. At each point, both the correction factor $k$ in region $B$ and the total time $T$ have been varied to maximize the fidelity.}
	\label{kappa}
\end{figure}

\begin{figure}[b]
	\includegraphics[width=0.5\textwidth]{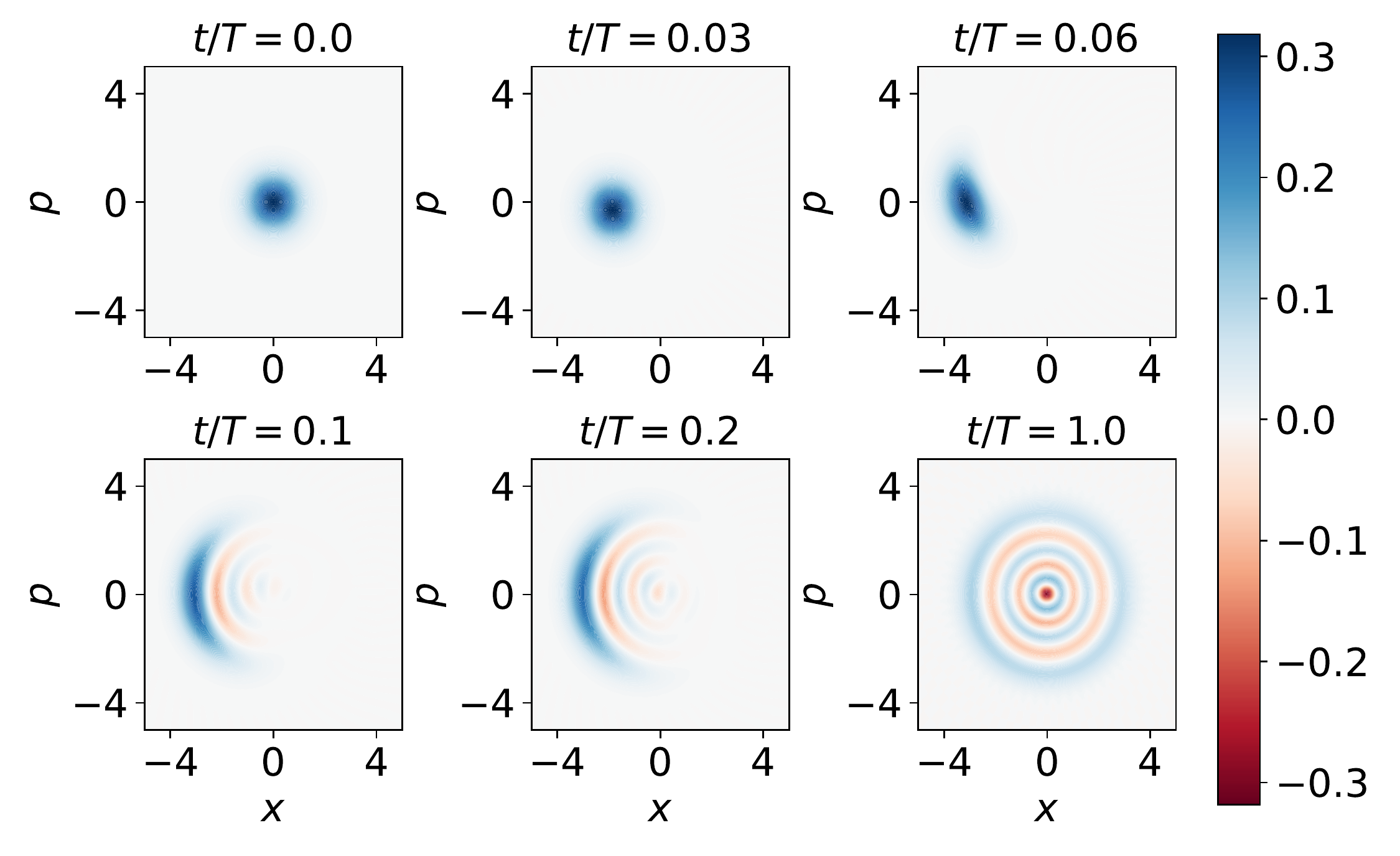}
	\caption{Wigner functions of the state at various times $t$ in the process of generating $\ket{5}$ with total time $T=11/\chi$ and decoherence rate $\kappa=\chi/1000$. The states in regions A ($t/T=0.0, 0.03$) and B ($t/T=0.06$) are well approximated as coherent states in gradually increasing displacements. In region C ($t/T=0.1, 0.2, 1.0$), the photon number uncertainty is gradually reduced to reach a Fock state by adiabatic passage through displaced Fock states.}
	\label{wigner}
\end{figure}

We simulate the master equation \eqref{master} for the open system using the optimized path for the corresponding closed system, which should establish reasonable estimates for the performance of our scheme in the presence of dissipation. In these simulations, we focus on the fidelities $F=\left<n\middle|\rho(T)\middle| n\right>$ of the generated states relative to the desired Fock states. The results are shown in Fig.~\ref{kappa}, and in Fig.~\ref{wigner}, we show intermediate states that appear in the process of generating $\ket{5}$ with $\chi/\kappa=1000$.

Empirically, we observe Rabi oscillations between $\ket{\phi_0}$ and $\ket{\phi_1}$ in region B, which can produce oscillations in the fidelity. To address these oscillations, we can extend the time spent in region B by a factor of $k$, which allows us to fine tune the phase of the Rabi oscillation at the end of region B to maximize the fidelity. We perform simulations for various $T$ and $k$ and take the highest $F$.

In considering these numerical results, it is worth mentioning that nonlinearities of $\chi/\kappa>1000$ have already been demonstrated in microwave superconducting circuits~\cite{Peterer2015}, indicating that an experimental demonstration of our scheme is  feasible in the present state of the art, potentially establishing it as a convenient and simple way of generating intracavity Fock states.

As an extension of our method, we can replace $\beta(a+a^\dagger)$ in \eqref{hamiltonian} with a two-photon driving term ${p}\bigl({a^\dagger}^2+a^2\bigr)/2$ which leads to the ``Kerr parametric oscillator" Hamiltonian
\begin{equation}
H=\frac{\chi}{2}{a^\dagger}^2a^2+\Delta a^\dagger a+\frac{p}{2}\left({a^\dagger}^2+a^2\right),
\end{equation}
as introduced in Ref.~\cite{Goto2016a,Puri2017} for superconducting circuit architectures. As the two-photon term mediates a direct two-photon interaction between $\ket{n}$ and $\ket{n+2}$, the energy gap between these states scales as $p$, as opposed to $\beta^2$ in the original case. Therefore, in this system, we expect that even photon number Fock states $\ket{2n}$ can be generated more efficiently. In combination with the coherent drive, all Fock states are accessible. 


\begin{acknowledgments}
The authors wish to thank L.G.\ Wright for helpful discussions.

R.Y., E.N., T.O., and H.M.\ acknowledge funding from NSF award PHY-1648807. R.Y.\ also acknowledges funding from the Masason Foundation.
\end{acknowledgments}

\bibliography{mybib}

\end{document}